# Influence of chemical environment on the transition of alternating current electroosmotic flow

Yu Han [1,#], Zhongyan Hu [1,#], Kaige Wang [1] and Wei Zhao [1,*]

[1] State Key Laboratory of Photon-Technology in Western China Energy, International Scientific and Technological Cooperation Base of Photoelectric Technology and Functional Materials and Application, Laboratory of Optoelectronic Technology of Shaanxi Province, Institute of Photonics and Photon-Technology, Northwest University, Xi'an, 710127, China
# The authors contributed equally to this investigation
* Correspondence: zwbayern@nwu.edu.cn

**Abstract:** Electroosmotic flow (EOF) is a ubiquitous phenomenon at the solid-liquid interface when an external electric field is applied. Despite its prevalence, the characteristics and mechanisms of EOF driven by an alternating current (AC) electric field, particularly within complex chemical environments, have remained insufficiently understood, owing primarily to a scarcity of experimental data. In this investigation, we advance the comprehension of AC EOF by employing a high-resolution measurement technique — laser-induced fluorescent photobleaching anemometer (LIFPA). This method allows for precise empirical characterization of transient velocity of EOF along the electric double layer (EDL) far from electrode surfaces. We have discerned a distinct transition in AC EOF behavior — from linear to nonlinear — across a wide parameter space, such as the velocity of bulk flow, the AC electric field's frequency and intensity, and the pH of the bulk fluid. Moreover, the transition within the AC EOF is quantified by the transitional electric field intensity, $E_{A,C}$, paired with a correlated dimensionless parameter, $Z_{nlc}$. A power-law relationship between the linear term coefficient $Z_l$ and $Z_{nlc}$ has been established, with the scaling exponents determined by the pH value of the electrolyte solution. With these findings, we aspire not only to deepen the understanding of AC EOF transitions but also to establish a robust model that elucidates the interplay between the electric field and fluid flow in both linear and nonlinear regimes. This research potentially paves the way for more predictable and controllable electrokinetic processes in numerous applications, including micro-/nanofluidic systems, electrochemical reactions, and beyond.

## I. INTRODUCTION

In the past two decades, electroosmotic flows (EOFs) driven by temporally varying alternating current (AC) electric fields have been commonly used in various engineering applications such as biomedical engineering [1-5], micro/nanofluidics [6-10], electrochemistry systems [11-14], and the energy industry [15-17].

Fig. 1 illustrates the simplest electrokinetic (EK) flow model, which consists of two electrodes and a long straight channel with insulated walls. Two types of electric structures, known as the electric double layer (EDL, i.e., Stern-Diffuse layer) or electric triple layer (ETL, i.e., Stern-Diffuse-Diffusion layer), are formed on the surfaces of the insulating walls and electrodes, respectively.

When a low-amplitude AC electric field is applied, the EDLs on the electrode surfaces maintain equilibrium. The internal electric field in the bulk fluids responds linearly to the external AC electric field, resulting in a linear oscillating electroosmotic flow. This has been extensively studied both through numerical simulations and experimental investigations for both Newtonian [18-22] and non-Newtonian fluids [23].

When the strength of the electric field exceeds a certain threshold, nonlinear effects attributable to the ion concentration polarization (ICP) or faradaic ICP [24] at the electrode surfaces, nonzero net charge distribution, nonuniform internal electric field, and their interplay with fluid, emerge and give rise to a host of nonlinear flow phenomena. Early predictions by Rubinstein and Zaltzman [25, 26] suggested that vortical flow could result from the Rubinstein-Zaltzman instability. Further studies [27, 28] delved into the nonlinear EOF, distinguishing it into weakly and highly nonlinear regimes. Bridging to a broader context, both the minor and substantial nonlinear electric fields in the EDL of electrodes can provoke a nonlinear electric field response in the bulk fluid, which in turn may induce nonlinear EOF [29, 30] against insulated walls. The study of the intrinsic dynamics of AC EOF has largely been dependent on numerical simulations and theoretical analyses. Experimental investigations, particularly those capturing the instantaneous flow dynamics, are somewhat scarce and typically indirect. Santiago et al. [31] employed micro-particle image velocimetry (μPIV) to study the average velocity of AC EOF at a low Reynolds number of $3 \times 10^{-4}$. Wu et al. [32] utilized μPIV for examining low-frequency AC EOF through phase-averaging techniques. Nonetheless, the spatial resolution limitations of μPIV hinder precise measurements of fluid velocity fluctuations at or near the electric double layer (EDL). Moreover, the use of fluorescent particles in μPIV to measure the velocity field encounters the challenge of particle hysteresis, which complicates the accurate measurement of high-frequency and significant velocity fluctuations.

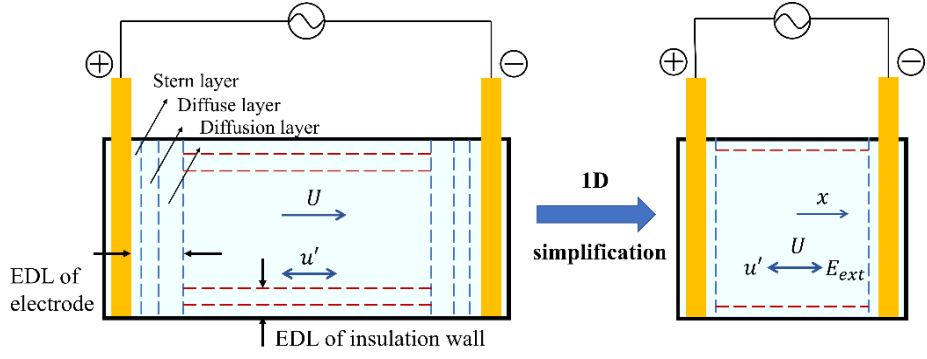

**Fig. 1.** Schematic and the 1D approximation of the AC EOF model.

While several experimental studies [33-36] have shed light on the behavior of EOF in its strongly nonlinear regime, our understanding about the transition from linear to nonlinear AC EOF is still insufficient. The cutting-edge laser-induced fluorescent photobleaching anemometry (LIFPA) [37] technique has made some progress towards understanding AC EOF. Zhao et al. have harnessed LIFPA to successfully capture the instantaneous velocity fluctuations of linear [21] and nonlinear [38-40] AC EOF adjacent to the EDL on the insulated wall. Notably, in linear AC EOF, they found that the mean square value of the velocity fluctuation ($u_{rms}$) diminishes proportionately to the frequency ($f_f$) raised to the power of $-0.66$, a finding that deviates from prior theoretical predictions [22] and suggests a slower response of AC EOF to the applied AC electric field than anticipated.

Further exploration into nonlinear AC EOF has identified the potential for varying states of flow, including periodic, quasiperiodic, or chaotic patterns [39], influenced by the basic flow rate, frequency, and intensity of the applied AC electric field. In a recent study conducted in 2022, Hu et al. [40] provided a comprehensive experimental analysis of the AC EOF transition from linear to nonlinear regimes. Their results pinpoint the transition electric field through the appearance of a second harmonic frequency within the power spectra of velocity fluctuations. These studies unveiled a power-law relationship between the dimensionless control parameters $Z_l$, which determines the influence of linear behavior, and $Z_{nlc}$, which signifies the onset of significant nonlinear behavior. Despite these efforts, the impact of pH levels on AC EOF dynamics remains unexplored.

In this paper, as a subsequent investigation of Hu et al. [40], we aim to undertake a systematic experimental study of the transition of AC EOF from linear to weakly nonlinear states, with a particular focus on the impact of pH value. Building on the existing theoretical framework, we have refined our analytical approach, leading to the development of new control equations tailored to capture the dynamic of AC EOF. Through systematically investigation, we expect to provide a more holistic picture of EOF behavior, which is crucial for the design and optimization of microfluidic and electrokinetic systems where precise control of flow is important.

## II. EXPERIMENTAL SETUP

### A. LIFPA developed on a confocal microscope

LIFPA is a novel micro/nanofluidics velocity measurement technique based on laser induced fluorescence (LIF) and photobleaching effect. [37] It offers ultrahigh spatial and temporal resolutions while maintaining electric neutrality, making it suitable for studying transient EK flow, e.g. characterizing EK turbulent flow in a micromixer [41-44], as well as linear/nonlinear EOF on the insulated wall [21, 38-40].

The LIFPA system used in this investigation is consistent to Hu et al [40], as shown in Fig. 2(a). Its lateral and axial spatial resolutions are about 180 nm and 800 nm [40], respectively. The details have been introduced in the supporting information.

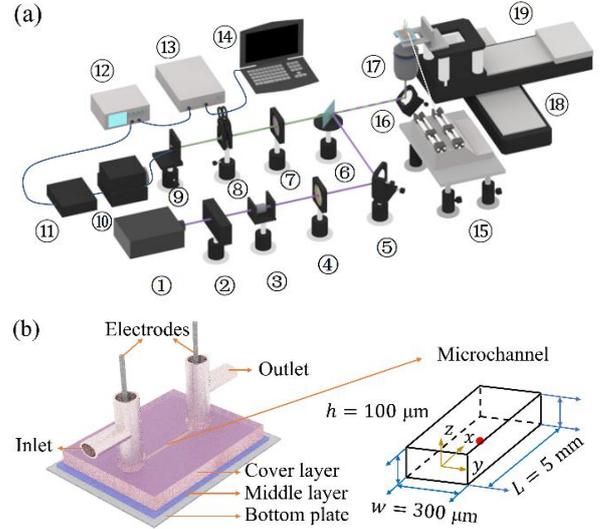

**Fig. 2.** (a) Schematic of the LIFPA system, including (1) 405 nm CW laser; (2) acoustooptic modulator; (3) spatial filter; (4) lens; (5) reflect mirror; (6) dichroic mirror; (7) lens;(8) band-pass filter; (9) band-pass filter and optical fiber connector; (10) PMT; (11) current preamplifier and A/D converter; (12) arbitrary function generator; (13) high voltage amplifier; (14) computer; (15) syringe pump; (16) reflect mirror; (17) objective lens; (18) micrometer translation stage; (19) nanometer piezo translation stage. (b) The schematic of the microchip and the coordinate system and the dimensions of the microchannel. The detection position is right at the center of the microchannel as depicted by the red spot.

**B. Working fluids**

The experiment used 100 μM coumarin 102 (Sigma Aldrich, St. Louis, MO, USA) aqueous solutions with varying electric conductivity and pH values. The solution was prepared with 5% methanol, which is used to dissolve coumarin 102 powder. The thickness of the EDL on the bottom of the channel was evaluated using Debye length $\lambda = \sqrt{\varepsilon k_B T / N_A e^2 \sum_i c_i z_i^2}$, with $\varepsilon = \varepsilon_0 \varepsilon_r$ being the electric permittivity. $\varepsilon_r$ is the relative electric permittivity of water and $\varepsilon_0$ is vacuum permittivity. $k_B$ is Boltzmann constant, $T$ is solution temperature, $N_A$ is Avogadro constant, $e$ is elementary charge, $c_i$ and $z_i$ are the concentration and electric valence of the ith ion. The values for each parameter are presented in Table SI.

**C. Microchip and experimental system**

Fig. 2(b) illustrates the microchip used in this investigation, which contains a 5 mm long ($L$), 300 μm wide ($w$), and 100 μm high ($h$) straight microchannel. The microchip was assembled using a layer-by-layer process [45] and comprises three layers, including cover layer, intermediate layer and bottom plate. Two platinum electrodes with a diameter of 100 μm are mounted at the inlet and outlet of the microchannel to generate AC EOF. The AC electric field is supplied by an arbitrary function generator (Tektronix AFG3102C) accompanied by a high voltage amplifier (Trek model PZD700A). In this investigation, the influence of basic flow velocity is taken into account, provided by a syringe pump (HARVARD PUMP 33).

**D. Velocity calibration of LIFPA**

LIFPA quantifies flow velocity by utilizing the photobleaching effect of fluorescence. In each experimental set, no matter the chemical conditions are changed or not, a velocity calibration curve showing the monotonic relationship between the fluorescent intensity ($I_f$) and the flow velocity ($U$) has been acquired, as plotted in Fig. S1. Please note, in different experimental set, the velocity calibration curve is different. The velocity calibration curve is nonlinearly fitted using a fifth-order polynomial function.

## III. ONE-DIMENSIONAL OSCILLATING EOF

In this investigation, the status of AC EOF on the electrodes is indirectly characterized by the velocity fluctuations of the oscillating EOF induced on insulated wall. This problem can be described generally by Poisson-Nernst-Planck-Navier-Stokes (PNPNS) equations [38, 40] as

$$\rho_f \left( \frac{\partial \vec{u}}{\partial t} + \vec{u} \cdot \nabla \vec{u} \right) = -\nabla p + \mu \nabla^2 \vec{u} + \rho_e \vec{E} \quad (1)$$

$$\frac{\partial \rho_i}{\partial t} + \vec{u} \cdot \nabla \rho_i = \nabla \cdot b_i z_i \rho_i \vec{E} + D_i \nabla^2 \rho_i \quad (2)$$

$$\rho_e = \varepsilon \nabla \cdot \vec{E} \quad (3)$$

where $\vec{u} = u\vec{x} + v\vec{y} + w\vec{z}$ is the velocity vector, $u$, $v$ and $w$ represent the magnitude of velocity in $x$, $y$ and $z$ directions. $\rho_f$ is the density of fluid, $P$ is pressure, $\mu$ is dynamic viscosity, $\varepsilon$ is electric permittivity of fluid, $\rho_e = \sum_i \rho_i z_i e$ is net charge density, with $\rho_i = c_i N_A$, $b_i$, $z_i$ and $D_i$ being the density, mobility, valence and diffusivity of the i$^{th}$ ion, respectively. $\vec{E} = E_x\vec{x} + E_y\vec{y} + E_z\vec{z}$ is the electric field vector, with $E_x$, $E_y$ and $E_z$ being the electric field components in $x$, $y$ and $z$ directions, respectively. Besides, considering the fluid is incompressible, $\nabla \cdot \vec{u} = 0$.

In this study, the focus is on the transition of the AC EOF from a linear to a nonlinear state. The oscillating AC EOF is generated near the EDL formed on the insulation wall, and is driven by the internal electric field, which is generated by the external electric field applied to the electrodes in the $x$-direction. Both the fluid movement and the electric field are parallel to the $x$-axis, which means that $v = w = 0$ and $\partial u / \partial x = 0$ according to incompressibility.

A one-dimensional (1D) approximation [38, 40] of the electric field has been utilized in this model. Eqs. (1)-(3) under 1D electric field approximation become

$$\rho_f \frac{\partial u'}{\partial t} = -\frac{\partial (\bar{p} + p')}{\partial x} + \mu \frac{\partial^2 (\bar{u} + u')}{\partial y^2} + \rho_e E_x \quad (4)$$

$$\frac{\partial \rho_i'}{\partial t} + \bar{u} \frac{\partial}{\partial x} \rho_i' + u' \frac{\partial}{\partial x} \rho_i' =$$
$$b_i z_i \rho_{i0} \frac{\partial}{\partial x} E_x + b_i z_i \frac{\partial}{\partial x} \rho_i' E_x + D_i \frac{\partial^2}{\partial x^2} \rho_i' \quad (5)$$

$$\rho_e = \sum_i (\rho_{i0} + \rho_i') z_i e = \varepsilon \left( \frac{\partial E_x}{\partial x} + \frac{\partial E_y}{\partial y} \right) \quad (6)$$

where $u' = u - \bar{u}$ is the velocity fluctuation, with $\bar{u}$ being the mean flow velocity and ¯ represent temporal averaging. $p' = p - \bar{p}$ is the pressure fluctuation, with $\bar{p}$ being the mean pressure. $\rho_i' = \rho_i - \rho_{i0}$ is the fluctuation of ion density caused by external electric field, with $\rho_{i0}$ being the original density of different ions in the fluid and $\partial \rho_{i0} / \partial x = 0$.

The problem involves the interaction between pressure-driven flow, AC EOF, and ion transport. However, it can be broken down into two consequent problems. When pressure-driven flow is applied, it interacts with the external AC electric field, causing ions to be transported towards or away from the electrodes. This results in fluctuations in ionic concentration near the electrodes, especially when the external electric field is strong. These fluctuations, in turn, lead to a nonlinear response in the internal electric field. The resulting internal electric field then drives the EDL formed on the insulated wall, generating oscillating EOF away from the electrodes. Each of these mechanisms operates based on different physical principles, and thence dimensions, requiring coordination of pressure-driven flow and the applied electric field to describe the flow phenomena.

For the basic pressure-driven flow, it can be characterized by the dimensional analysis in the following

$$\bar{u} = \bar{u}^* U_p, \frac{\partial^2 \bar{u}}{\partial y^2} \sim U_p h^{-2} \frac{\partial^2 \bar{u}^*}{\partial y^{*2}}, \bar{p} = \bar{p}^* \rho_f U_p^2, \frac{\partial \bar{p}}{\partial x} \sim \frac{\rho_f U_p^2}{l_0} \frac{\partial \bar{p}^*}{\partial x^*} \quad (7)$$

where $U_p$ is the bulk flow velocity of the pressure-driven flow, $l_0$ is a large scale to characterize the quantitative variation in streamwise direction.

For the oscillating AC EOF, the physical quantities can be analyzed by the following dimensions

$$t = t^*/f_f, x = x^* l_0, y = y^* \lambda, u' = u'^* U_0, \frac{\partial^2 u'}{\partial y^2} \sim U_0 \lambda^{-2} \frac{\partial^2 u'^*}{\partial y^{*2}},$$

$$p' = p'^* \rho_f U_0^2, \rho_i' = \rho_i'^* \rho_{i0}, \rho_i = \rho_i^* \rho_{i0}, \rho_{e,EDL} = \frac{\varepsilon|\zeta|}{\lambda^2} \rho_{e,EDL}^*,$$

$$E_x = E_x^* E_A, \frac{\partial}{\partial x} E_x \sim \frac{E_A}{l_0} \frac{\partial}{\partial x^*} E_x^*, \frac{\partial}{\partial y} E_y \sim \frac{|\zeta|}{\lambda^2} \frac{\partial}{\partial y^*} E_y^* \quad (8)$$

where $U_0 = \varepsilon|\zeta|E_A/\mu$ [22, 46-49] is the characteristic velocity of oscillating AC EOF, $\rho_{e,EDL}$ is the electric charge density on the surface of electrode and insulated wall, respectively. $V_T = k_B T/ze$ is thermal potential. $\zeta$ is zeta potential on the insulated bottom. $E_A$ is the amplitude of the external electric field with $E_A = V_{ext}/l_0$ and $V_{ext}$ is the amplitude of the external AC voltage across the electrodes.

Accordingly, the dimensionless forms of Eqs. (4)-(6) become

$$\frac{\partial u'^*}{\partial t^*} = -\frac{U_p^2}{l_0 U_0 f_f} \frac{\partial \bar{p}^*}{\partial x^*} - \frac{U_0}{l_0 f_f} \frac{\partial p'^*}{\partial x^*} + \frac{\mu U_p}{h^2 U_0 f_f \rho_f} \frac{\partial^2 \bar{u}^*}{\partial y^{*2}}$$
$$+ 2\pi \frac{l_s^2}{\lambda^2} \frac{\partial^2 u'^*}{\partial y^{*2}} + \frac{\varepsilon|\zeta|E_A}{\rho_f \lambda^2 U_0 f_f} \rho_{e,EDL}^* E_x^* \quad (9)$$

$$\frac{\partial \rho_i'^*}{\partial t^*} + \frac{U_p}{l_0 f_f} \bar{u}^* \frac{\partial \rho_i'^*}{\partial x^*} + \frac{U_0}{l_0 f_f} u'^* \frac{\partial \rho_i'^*}{\partial x^*} =$$
$$\frac{b_i z_i E_A}{l_0 f_f} \frac{\partial E_x^*}{\partial x^*} + \frac{b_i z_i E_A}{l_0 f_f} \frac{\partial \rho_i'^* E_x^*}{\partial x^*} + \frac{D_i}{l_0^2 f_f} \frac{\partial^2 \rho_i'^*}{\partial x^{*2}} \quad (10)$$

$$\rho_e = \varepsilon \left( \frac{E_A}{l_0} \frac{\partial E_x^*}{\partial x^*} + \frac{|\zeta|}{\lambda^2} \frac{\partial E_y^*}{\partial y^*} \right) \quad (11)$$

with $\rho_e = \sum_i (\rho_{i0} + \rho_i')z_i e = \varepsilon \frac{E_A}{l_0} \frac{\partial E_x^*}{\partial x^*}$ and $\rho_{e,EDL} = \varepsilon \frac{|\zeta|}{\lambda^2} \frac{\partial E_y^*}{\partial y^*}$ are the net charge density in bulk region and EDL, respectively. For the sake of simplicity, we consider a simple binary symmetric electrolyte. The positive and negative ions have equivalent magnitude of valance ($z_+ = -z_- = z$), accordingly, $z_i^2 = z^2$ and $\rho_{+0} = \rho_{-0} = \rho_0$. We further assume the positive and negative have equivalent $D_i$ with $D_+ \approx D_- = D$. Considering Nernst-Einstein equation $D_i = b_i k_B T/z_i e$, it is obtained $b_+ \approx b_- = b$. Since the solution in bulk region is electrically neutral ($(\rho_{+0}' - \rho_{-0}')ze = 0$), we have $\rho_e = (\rho_+' - \rho_-')ze$. Thus, after simple processing on Eq. (10) and (11), it is obtained

$$\frac{\partial}{\partial t^*} \frac{\partial E_x^*}{\partial x^*} + Z_l \bar{u}^* \frac{\partial}{\partial x^*} \frac{\partial E_x^*}{\partial x^*} + Z_E u'^* \frac{\partial}{\partial x^*} \frac{\partial E_x^*}{\partial x^*} =$$
$$2 \frac{bz^2 e \rho_0}{\varepsilon f_f} \frac{\partial E_x^*}{\partial x^*} + \frac{bz^2 e \rho_0}{\varepsilon f_f} \frac{\partial (\rho_+'^* + \rho_-'^*)E_x^*}{\partial x^*} + \frac{bV_T}{l_0^2 f_f} \frac{\partial^2}{\partial x^{*2}} \frac{\partial E_x^*}{\partial x^*} \quad (12)$$

Here, we define $Z_l = U_p/l_0 f_f$ (a reciprocal of Strouhal number), $Z_E = U_0/l_0 f_f$ for convenience.

Our objective is to investigate the transition of AC EOF on the electrodes from a linear to a nonlinear state. In this context, the transition electric field refers to the minimum electric field required for the initiation of nonlinear AC EOF flow, specifically in the weakly nonlinear region. In this region, the fluctuation in ion density is extremely small and can be neglected, meaning that the ion density can be approximated as $\rho_+'^* \approx 0$. Thus, $\frac{\partial}{\partial x^*}(\rho_+'^* + \rho_-'^*)E_x^* = \frac{\partial}{\partial x^*}\left(2\rho_+'^* - \frac{\varepsilon E_A}{ze\rho_0 l_0} \frac{\partial E_x^*}{\partial x^*}\right)E_x^* \approx -\frac{\partial}{\partial x^*}\left(\frac{\varepsilon E_A}{ze\rho_0 l_0} \frac{\partial E_x^*}{\partial x^*}\right)E_x^*$. Besides, the velocity fluctuations due to AC EOF are much smaller than the pressure-driven flow, i.e. $U_p \gg U_0$ and $Z_l \gg Z_E$. Based on these approximations and integrating over $x^*$, Eq. (12) becomes

$$\frac{\partial E_x^*}{\partial t^*} + Z_l \bar{u}^* \frac{\partial E_x^*}{\partial x^*} - Z_f E_x^* + Z_{nl} E_x^* \frac{\partial E_x^*}{\partial x^*}$$
$$= Z_d \frac{\partial^2 E_x^*}{\partial x^{*2}} + A(t) \quad (13)$$

where $A(t)$ is an temporal function determined by the applied external electric field, $Z_{nl} = bzE_A/l_0 f_f$ represents the influence of nonlinear electric field, $Z_f = 2bz^2 e\rho_0/\varepsilon f_f$ represents the influence of electrophoresis and $Z_d = bV_T/l_0^2 f_f = D/l_0^2 f_f = l_{od}^2/l_0^2$ (where $l_{od} = \sqrt{D/f_f}$ is an oscillating diffusion scale of ions, representing the diffusion length in a forcing period) represents the influence of diffusion. $Z_d$ can be important when $l_0 \to 0$, e.g. when the electrodes are sufficient close to each other. However, in this study, since $l_0 = L$ is the distance between two electrodes, we have $Z_f \gg Z_l \gg Z_{nl} \gg Z_d$, therefore, the diffusion term can be neglected. Eq. (13) becomes

$$\frac{\partial E_x^*}{\partial t^*} + Z_l \bar{u}^* \frac{\partial E_x^*}{\partial x^*} + Z_{nl} E_x^* \frac{\partial E_x^*}{\partial x^*} = Z_f E_x^* + A(t) \quad (14)$$

Thus, after dimensional analysis, the Nernst-Planck equation (Eq. (10)) is turned into Eq. (14) which is a correction of the one by Hu et al [38].

From Eq. (9), we further get the perturbation equation of velocity after a simple process, as

$$\frac{\partial u'^*}{\partial t^*} - 2\pi \frac{l_s^2}{\lambda^2} \frac{\partial^2 u'^*}{\partial y^{*2}} = \frac{\varepsilon|\zeta|E_A}{\rho_f \lambda^2 U_0 f_f} \rho_{e,EDL}^* E_x^* \quad (15)$$

For an oscillating EOF driven by AC electric field, as inferred by the theoretical analysis of Dutta and Beskok [50] and the experimental investigations of Zhao et al [21], $u'_{EDL} = u'(y = \lambda)$ is also determined by AC frequency, say

$$u'_{EDL} \sim \frac{\varepsilon |\zeta| E}{\mu} G(f_f) \quad (16)$$

The dimensionless function $G(f_f)$ is a correction factor that will be determined empirically in the following. Note, $E$ is the local and internal electric field that carries either linearity or nonlinearity corresponding to external electric field. Eq. (16) can be nondimensionalized as $u'^*_{EDL} = E^*_x G(f_f)$ on the EDL of the insulated wall. After substituting it into Eq. (15), it is obtained

$$\frac{\partial u'^*_{EDL}}{\partial t^*} + Z_l \bar{u}^* \frac{\partial u'^*_{EDL}}{\partial x^*} + Z_{nl} G^{-1}(f_f) u'^*_{EDL} \frac{\partial u'^*_{EDL}}{\partial x^*}$$
$$= Z_f u'^*_{EDL} + G(f_f) A(t) \quad (17)$$

In this study, by measuring $u'_{EDL}$, the influence of $Z_l$ and $Z_{nl}$ on the status of AC EOF can be revealed.

## IV. EXPERIMENTAL RESULTS

### A. Statistics of the velocity field

#### 1) Mean velocity

The mean velocity $\bar{u}$ at different frequencies and different flow rates $Q$ is measured first, as shown in Fig. 3. In the unforced case where $f_f = 0$ Hz, $\bar{u}$ varies with $Q$ in an approximately linear manner as expected. When the AC electric field is applied, $\bar{u}$ at different AC frequencies only exhibit negligibly small derivations from the unforced case. This indicates that within the considered range of frequency and intensity of the AC electric field, the applied AC electric field has little effect on the mean velocity of the flow. In other words, no noticeable mean flow is induced by the AC EOF.

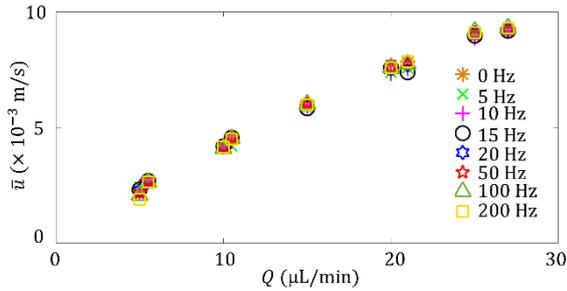

Fig. 3. Relationship between flow rate $Q$ and mean velocity $\bar{u}$ at different AC frequencies, where $E_A = 8 \times 10^4$ V/m. Here, $\sigma = 6.08$ μS/cm and pH = 7.11. The sampling rate is 1 kHz.

#### 2) Influence of the frequency of AC electric field

This section aims to demonstrate how velocity fluctuations, expressed as $u_{rms}/U_0$—with $u_{rms} = \sqrt{\overline{u'^2}}$ being the root mean square of velocity fluctuations—vary under different control conditions as shown in Figs. 4(a, b). A remarkable observation is that, regardless of whether $E_A$ is small or large, $u_{rms}/U_0$ is directly proportional to $\ln Z_l$ as following

$$\frac{u_{rms}}{U_0} \sim A - \beta \ln Z_l \quad (18)$$

where $A$ is an initial value of $u_{rms}/U_0$ corresponding to flow rate $Q$. $\beta < 0$ is a scaling exponent as elucidated below. Considering $U_0 = \varepsilon |\zeta| E_A / \mu$ and $Z_l = U_p / l_0 f_f$, Eq. (18) can be rewritten as

$$\frac{u_{rms}}{U_0} \sim \ln e^A \left(\frac{f_f}{f_r}\right)^\beta \quad (19)$$

where $f_r = U_p/l_0$. Through nonlinear fitting of Eq. (19), it was discovered that for $E_A$ values of $5 \times 10^3$, $2 \times 10^4$ and $E_A = 8 \times 10^4$ V/m, $\beta$ is approximately $-0.58$, $-0.6$, and $-0.66$ in average, respectively. This suggests that there is a greater $\beta$ value associated with higher $E_A$, indicating a stronger nonlinearity. In other words, a more negative $\beta$ is present when there is a stronger nonlinearity. Furthermore, when $\ln e^A (f_f/f_r)^\beta$ is approximated to $e^A (f_f/f_r)^\beta$ after Taylor expansion, it is found that the $\beta$ value for linear AC EOF, which is $-0.58$, is approximately consistent with the previously reported value of $-0.66$ [21].

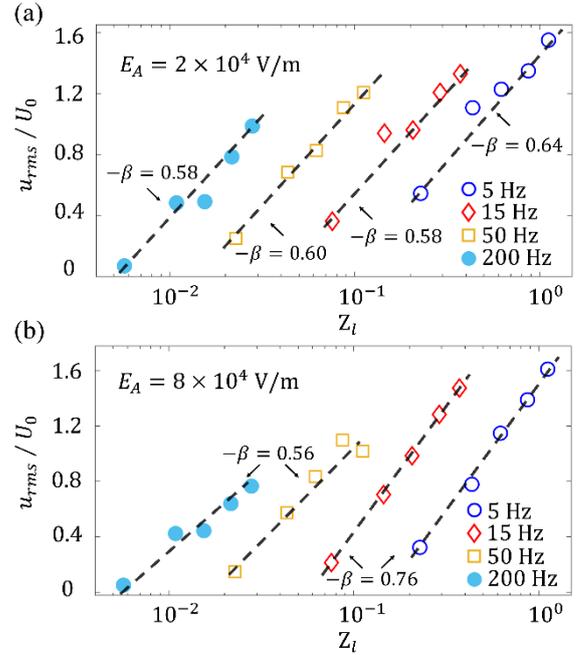

Fig. 4. Relationship between $u_{rms}/U_0$ and $Z_l$ at different $Q$ and $f_f$. The sampling rate is 1 kHz. Here, $\sigma = 6.08$ μS/cm and pH = 7.11. (a) $E_A = 2 \times 10^4$ V/m; (b) $E_A = 8 \times 10^4$ V/m.

### B. Velocity power spectrum

The fundamental distinction between linear and nonlinear AC EOF lies in the respective responses of the internal electric field and corresponding EOF velocity to an external AC electric field. Specifically, while linear AC EOF responds proportionally to external electric field, nonlinear AC EOF exhibits a nonlinear response–including distortion–as predicted by Olesen et al. [28]. This means that nonlinear AC EOF's velocity power spectrum may differ significantly from that of the external AC electric field. To provide further insight into the evolution of AC EOF, the velocity power spectrum ($S(f)$) of $u'$ is analyzed in accordance with the criterion established by Hu et al. [40] to differentiate linear and nonlinear regimes (e.g. as demonstrated in Fig. 5(a)). To facilitate interpretation, the results are presented with respect to dimensionless frequency $f^* = f/f_f$.

1) $S(f)$ varies with $E_A$

In Fig. 5(a), a sinusoidal AC signal with a frequency of $f_f = 5$ Hz is applied to the system. The $E_A$ gradually increases from $5 \times 10^3$ to $2 \times 10^4$ V/m. At $E_A = 5 \times 10^3$ V/m, only a single peak is observed in the velocity power spectrum at $f^* = 1$, indicating that the EOF behavior is linear in response to the applied AC electric field. When $E_A$ is increased to $1 \times 10^4$ V/m, two peaks appear in the power spectrum at $f^* = 1$ and 2. The appearance of high-order harmonics at $f^* = 2$ indicates that the AC EOF has entered a nonlinear regime. As $E_A$ increases, the peak value at $f^* = 1$ reaches saturation at $E_A = 4 \times 10^4$ V/m, as observed in Fig. 5(b). Meanwhile, the peak values at higher-order harmonic frequencies increase rapidly, indicating increasing nonlinearity. Although the peak values generally decrease with $f^*$, an interesting phenomenon occurs occasionally: the peak values at odd $f^*$ can be even higher than those at neighboring even $f^*$. For instance, when $E_A = 2 \times 10^4$ V/m, $f^* = 3$ exhibits a larger peak value than $f^* = 2$. This may be attributed to a resonant interaction between the AC electric field and the RC dynamics, which has a characteristic RC frequency of approximately $f_{RC} \approx 5$ Hz [27], half of $f_f$ in this case. The growth of peak values at odd-order harmonic frequencies is significantly enhanced, while those of even-order frequencies are slightly inhibited.

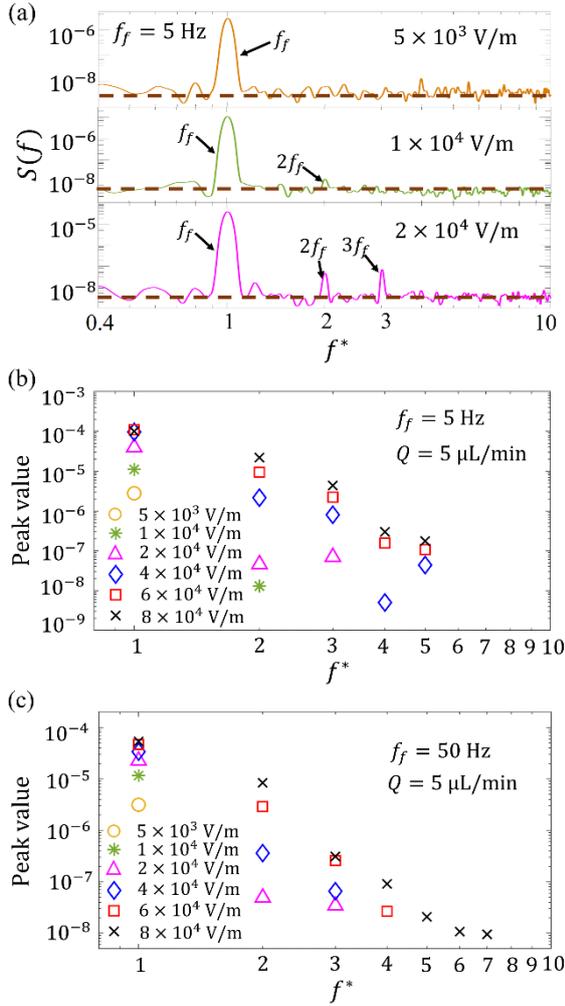

When $f_f = 50$ Hz (Fig. 5(c)), the frequency of the AC electric field is much greater than $f_{RC}$. The resonant interaction mechanism between the AC electric field and the RC dynamics is not significant. Therefore, in this scenario, we do not observe the phenomenon where the spectral values at odd $f^*$ increase faster compared to even $f^*$.

2) $S(f)$ of $u'$ varies with $Q$

The influence of $Q$ on the status of AC EOF can be observed directly in the spectral domain. For instance, at $E_A = 2 \times 10^4$ V/m (as shown in Fig. 6), $S(f)$ exhibits peaks at $f^* = 1, 2,$ and 3, indicating the presence of nonlinear behavior. However, as $Q$ is increased to 10 µL/min, only two peaks persist, and with further increase to 27 µL/min, only a single peak at $f^* = 1$ can be observed. The peaks at higher harmonic frequencies vanish, showing the significance of the linear effect ($Z_l$) becomes more prominent.

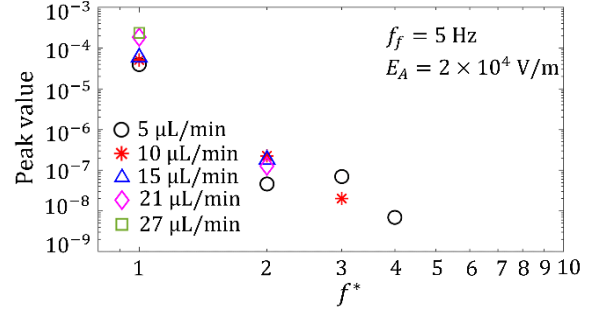

**Fig. 6.** Peak values of $S(f)$ at different $f^*$. The sampling rate is 1 kHz. Here, $\sigma = 6.08$ µS/cm, pH = 7.11, $f_f = 5$ Hz, and $E_A = 2 \times 10^4$ V/m.

3) $S(f)$ of $u'$ varies with $f_f$

The suppression of nonlinearity can also be realized by increasing $f_f$. Specifically, when $E_A = 5 \times 10^3$ V/m, the electric field is not sufficiently high to induce nonlinear AC EOF for $f_f \in [5, 200]$ Hz. In this case, there is only a single peak at $f^* = 1$ for all $f_f$, as depicted in Fig. 7(a). According to the results of the nonlinear fitting, the peak value is proportional to $\ln f_f^{-0.58}$, with $\beta = -0.58$. This is consistent to the result in section 3.1. In contrast, at $E_A = 8 \times 10^4$ V/m (Fig. 7(b)), as $f_f$ is increased, the peak values of the higher-order harmonic frequencies gradually decrease. Ultimately, the peaks fall below the noise level.

**Fig. 5.** (a) Power spectra of $u'$ under different $E_A$, with $f_f = 5$ Hz. The sampling rate is 1 kHz. Here, $\sigma = 6.08$ µS/cm, pH = 7.11, $Q = 5$ µL/min. The dashed lines indicate the noise level of unforced flow. (b-c) Peak values of $S(f)$ at different $f^*$. (b) $f_f = 5$ Hz; (c) $f_f = 50$ Hz.

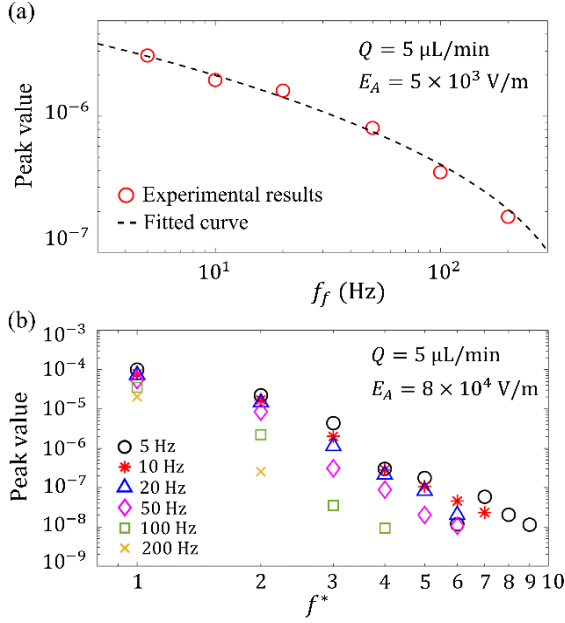

**Fig. 7.** Peak values of $S(f)$ at different $f^*$. The sampling rate is 1 kHz. Here, $\sigma = 6.08$ μS/cm, pH = 7.11, $Q = 5$ μL/min. (a) $E_A = 5 \times 10^3$ V/m, (b) $E_A = 8 \times 10^4$ V/m.

### C. Critical dimensionless parameters for different pH environments

Generally speaking, the variations of AC EOF with respect to $E_A$, $f_f$, and $Q$, as analyzed by statistical methods, time-series analysis, and spectral analysis, all exhibit consistent trends. These findings are coincident to the experimental observations reported by Hu et al. [40], lending support to their conclusions.

In the current section, we explore the influence of pH values and electric conductivities on the transitional electric field by examining the relationship between $Z_l$ and $Z_{nlc}$ across a broad range of parameters (by altering chemical conditions), where $Z_{nlc} = bzE_{A,C}/l_0 f_f$ is the critical $Z_{nl}$ calculated from the critical electric field amplitude $E_{A,C}$. The conductivity and pH of the solution are experimentally adjusted by the addition of 1X phosphate buffered saline (PBS) solution (HyClone, SH30256.01, Logan, UT, USA).

The experimental results are indicated by red dots in Fig. 8. As can be observed, at pH = 7.11 (Fig. 8(a)), there is a gradual increase in $Z_{nlc}$ alongside an increase in $Z_l$. When $Z_{nl}$ is greater than or equal to $Z_{nlc}$, the AC EOF enters a nonlinear regime; otherwise, it remains within a linear regime. At pH = 7.36 (Fig. 8(b)), within the same range of $Z_l$, $Z_{nlc}$ exhibits a slightly smaller slope compared to that at pH = 7.11. For instance, when $Z_l$ is at 1.12, $Z_{nlc}$ is just about $2.5 \times 10^{-2}$ at pH = 7.36, as opposed to $2.8 \times 10^{-2}$ at pH = 7.11. Moreover, as the pH is further incremented from 7.40 to 7.58 (Figs. 8(c, d)), the slope of the $Z_{nlc}$ curve continues to exhibit a decreasing trend. Conversely, when the pH value increases from 7.58 to 8.51 (Figs. 8(d-f)), the slope of the $Z_{nlc}$ curve starts to ascend once again, suggesting a higher difficulty for the AC EOF to attain transition. Despite some scatter in the data points in Fig. 8, they generally align with nonlinear curves that can be described by a power-law relationship, as follows:

$$Z_{nlc} = aZ_l^b + c \qquad (20)$$

The fitted results are depicted in Fig. 8 with black lines. The consistency between the experimental results and the fitted curves is pronounced. Below the $Z_{nlc}$ curve, the AC EOF exhibits predominantly linear behavior. In contrast, above the $Z_{nlc}$ curve, the nonlinear behavior of the AC EOF becomes significant.

As observed in Fig. 8, the curvature of the $Z_{nlc}$ curve varies with the pH values. This curvature, quantified by the scaling index $b$, is directly correlated with the pH of the solution via the $\zeta$ potential. Behrens and Grier [51] have theoretically demonstrated that $\zeta \sim pH - pK$, where pK = 7.5 [52] is the dissociation constant. By employing a simple linear relationship $b = b_0 + b_1|\zeta| = b_0 + b_1|pH - pK|$, the relationship between $b$ and pH can be approximately described by the following fit

$$b = 0.66 + 0.5 \times |pH - pK| \qquad (21)$$

The plot presented in Fig. 9 qualitatively supports the fitting obtained using the linear model. This is also consistent to the investigation of Hu et al [40].

## V. CONCLUSIONS

In this study, AC electroosmotic flow is thoroughly examined both theoretically and experimentally, focusing on the dynamics of flow velocity within an expansive parameter space, such as bulk flow velocity, the frequency and magnitude of the AC electric field, and notably, the pH level of the bulk fluid. A one-dimensional simplification facilitates the derivation of control equations that govern AC EOF, impacting both the velocity field and the internal electric field.

The behavior of AC EOF can be discerned by evaluating two coefficients: the linear term coefficient ($Z_l$) and the nonlinear term coefficient ($Z_{nl}$). On one hand, a large $Z_l$ value represents the significance of the linear term within the control equation, which in turn means that the AC EOF is more likely to exhibit a linear response to the applied electric field. This effect can be achieved by increasing the basic flow rate and the AC frequency. We observe that the velocity fluctuations correspond to $\ln f_f^\beta$, where $\beta$ is $-0.58$. This variation aligns with findings from past studies [21]. On the other hand, an augmented $Z_{nl}$ emphasizes the role of the nonlinear term in the control equation, which arises under conditions of either an intensified electric field or a reduced AC frequency. Increasing nonlinearity in AC EOF correlates with a more substantial negative $\beta$. Furthermore, during the evolution of a markedly nonlinear AC EOF, resistor-capacitor dynamics become crucial.

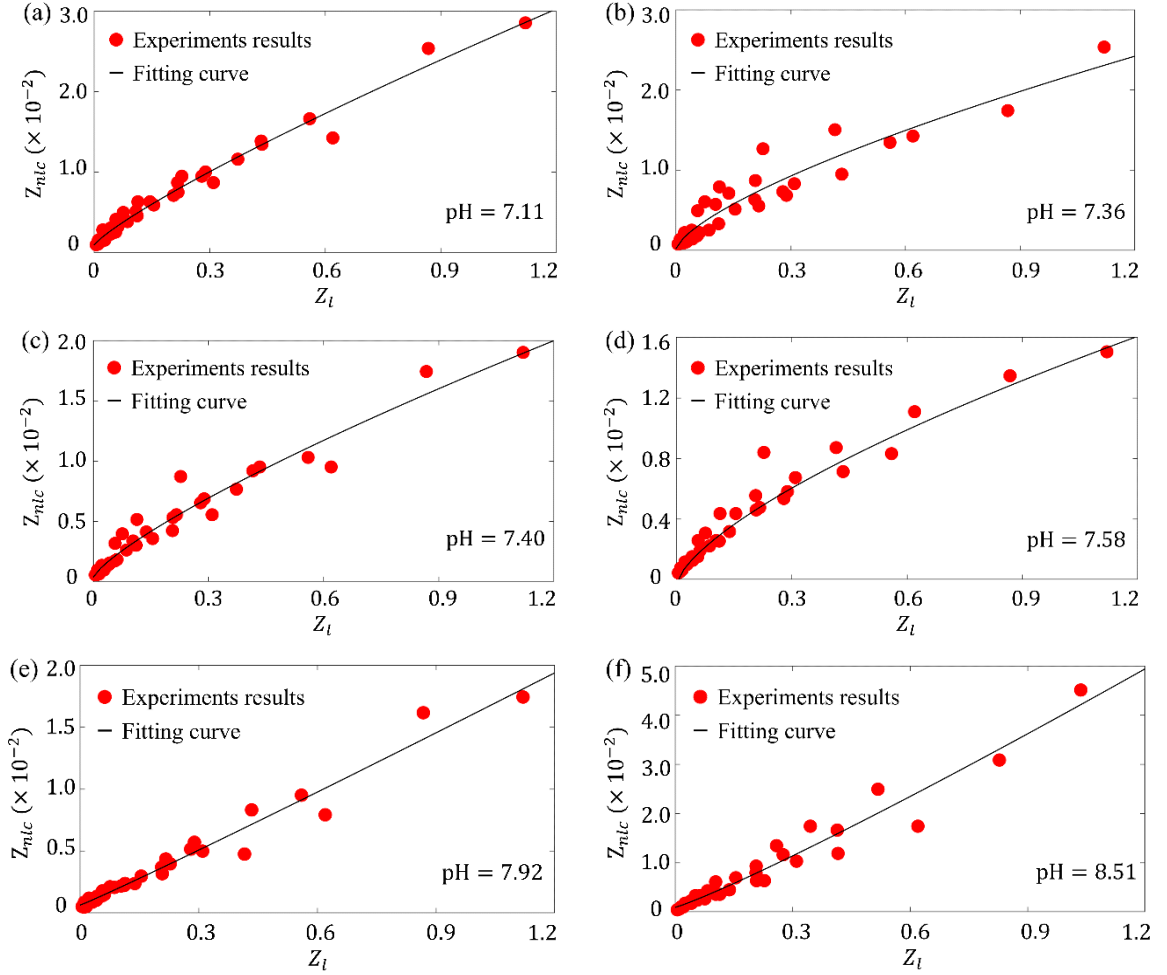

Fig. 8. Relationship between $Z_l$ and $Z_{nlc}$ in six solutions with different pH values. (a) Fluorescent solution without PBS, $\sigma = 6.08$ μS/cm, pH=7.11; (b) fluorescent solution with PBS, $\sigma = 10.43$ μS/cm, pH=7.36; (c) fluorescent solution with PBS, $\sigma = 14.89$ μS/cm, pH=7.40; (d) fluorescent solution with PBS, $\sigma = 25.6$ μS/cm, pH=7.58; (e) fluorescent solution with PBS, $\sigma = 41.6$ μS/cm, pH=7.92; (f) fluorescent solution with PBS, $\sigma = 86$ μS/cm, pH=8.51.

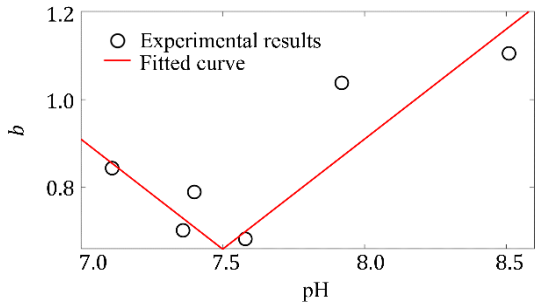

**Fig. 9.** Relationship between the scaling exponent $b$ and pH value.

Experimental observations aid in distinguishing between linear and nonlinear AC EOF regimes, delineated by the transitional electric field $E_{A,C}$ and the corresponding nonlinear coefficient $Z_{nlc}$, across various pH levels. Findings indicate that $Z_{nlc}$ shares a power-law relation with $Z_l$ within the experimental range, following $Z_{nlc} \sim Z_l^b$. Intriguingly, the scaling exponent $b$ is a function of the pH, described by the relationship $b = 0.66 + 0.5 \times |\text{pH} - \text{pK}|$. Consequently, solutions with a pH deviating considerably from the dissociation constant pK result in a large $b$, signifying a rapid and markedly nonlinear increase of $Z_{nlc}$ alongside $Z_l$. Under constant $Z_l$ conditions, a greater $Z_{nlc}$ is hence necessary for a larger $b$ when there is a greater deviation from pK, indicating that achieving nonlinearity in AC EOF becomes more challenging.

Our research aims to bridge gaps in experimental understanding of nonlinear AC EOF and elucidate the intricate interplay among AC EOF dynamics, pressure-driven basic flow, and the pH environment. Concurrently, these inquiries could hold substantial implications for practical application across micro-/nano-fluidic arenas, electric energy systems and devices, electrochemistry, and more. By judiciously selecting control parameters, such as $Z_l$ and $Z_{nl}$, it is possible to either promote or mitigate nonlinear EOF and finely tune fluid flow states. These capabilities enable either the augmentation of heat and mass transfer or the reduction of electrical energy waste, thereby enhancing energy efficiency.

## SUPPLEMENTARY MATERIAL

See the supplementary material for the LIFPA system, the property of the working fluids, velocity calibration curve, one-dimensional approximation of electric field and the mathematical processing from Eq. (10) to (12).

## ACKNOWLEDGEMENT

The authors appreciate the valuable suggestions from Guiren Wang. The investigation is supported by National Natural Science Foundation of China (51927804, 61378083, 61775181).

## AUTHOR DECLARATIONS

### Conflict of Interest

There are no conflicts of interest to declare.

### Author Contributions

**Yu Han:** Formal analysis (equal); Investigation (equal); Methodology (equal); Writing – original draft (equal). **Zhongyan Hu:** Formal analysis (equal); Investigation (equal); Methodology (equal); Writing – original draft (equal). **Kaige Wang:** Funding acquisition (equal); Methodology (equal); Resources (equal); Supervision (equal); Writing – review & editing (equal). **Wei Zhao:** Conceptualization (lead); Formal analysis (equal); Data curation (lead); Funding acquisition (equal); Resources (equal); Supervision (equal); Writing – review & editing (equal).

## DATA AVAILABILITY

The data that support the findings of this study are available from the corresponding authors upon reasonable request.

## NOTES AND REFERENCES